\documentstyle[prl,aps,epsfig,floats]{revtex}

\newcommand\beq{\begin{equation}}
\newcommand\eeq{\end{equation}}
\newcommand\bea{\begin{eqnarray}}
\newcommand\eea{\end{eqnarray}}

\newcommand\bi{\begin{itemize}}
\newcommand\ei{\end{itemize}}

\begin{document}

\draft

\textheight=23.8cm
\twocolumn[\hsize\textwidth\columnwidth\hsize\csname@twocolumnfalse\endcsname

\title{\Large Quantised charge pumping through multiple quantum dots}
\author{\bf Argha Banerjee$^1$, Sourin Das$^2$ and Sumathi Rao$^2$ } 

\address{\it $^1$
Department of Physics, Indian Institute of Technology, Kanpur,India}

\address{\it $^2$
Harish-Chandra Research Institute,
Chhatnag Road, Jhusi, Allahabad 211019, India}
\date{\today}
\maketitle

\begin{abstract}

We study electron pumping through a system of barriers, whose
heights are deformed adiabatically. We derive a simple
formula for the pumped charge $Q$ in terms of the total reflection
and transmission amplitudes and phases. 
The pumped charge increases with the number of 
barriers ($n_b$) and shows an
interesting step-like behaviour,
with the steps appearing at integer values of $Q$. 
The pumped charge also tends towards quantisation with the increase of 
the amplitude of the time-varying potential.
The value of the quantised pumped charge is 
shown to be correlated to the discontinuity 
of the reflection phase.

\end{abstract}


]

A parametric electron pump is a device which generates a dc current
at zero bias by cyclic deformations of the system 
parameters \cite{THOULESS,NIU}. The parameters of the Hamiltonian 
are slowly varied as a periodic function of time such that
the Hamiltonian returns to itself at the end of one cycle,
but charge has been pumped through the system.
In the last few years, electron pumps consisting of small semi-conductor
dots have received a great deal of experimental\cite{KOUW,POTHIER}
and theoretical attention\cite{SPIVAK,BROUWER,ALEINER,ZHOU,RENZONI}. 

Of more recent interest are the quantum pumps in open dot
systems, where quantum interference of the electronic wave-function
rather than Coulomb blockade (CB) is expected to play the
major role. Such a pump has been fabricated by Switkes 
{\it et al}\cite{SWITKES}. A scattering approach to such a parametric electron
pump was pioneered by Brouwer\cite{BROUWER} where 
the pumped current was related to parametric
derivatives of the scattering matrix. 
Using this, several theoretical papers\cite{WEI,ENTIN,ZHU}
have investigated the connection between resonant transmission
and the pumped charge. In general, the pumped charge  is
not quantised; however, it has been shown that when the
pumping contour encloses almost all of a resonance line, the
charge pumped is almost quantised\cite{ENTIN}.
It has also been shown that  inclusion of
inter-electron interactions using the
Luttinger liquid formalism\cite{SHARMA} leads to 
charge quantisation even with just a
double barrier system, due to the insulating
nature of the Luttinger wire in the presence of any barrier,

Although for the case of a rigidly sliding
potential $U(x-vt)$, where the periodicity in time $T$ is related
to the periodicity in space $L$ by $T=nL/v$, $n=$integer,
it can be proved using topological arguments that the charge is 
quantised\cite{THOULESS}, it is not so obvious that those arguments 
hold for more general cases where the Galilean principle does
not hold. Several explicit examples were considered by
Niu\cite{NIU} 
but quantisation has been shown only for potentials which
are spatially periodic.

In this letter, we derive a simple formula for the pumped charge
in terms of the reflection and transmission amplitudes of the system -
\beq
Q = {e\over 2\pi}\int_0^{\tau} dt ~{\dot \theta} 
-  {e\over 2\pi}\int_0^\tau dt ~t^2({\dot \theta}+{
\dot \phi})~,
\label{pcharge2}
\eeq
where $\tau$ is the time period of the perturbation
in the potential
which causes the charge to be pumped and $r,\theta$ and $t,\phi$
are the reflection and transmission amplitudes and phases respectively.   
We show that quantisation occurs whenever the contribution
from the second term vanishes. 

We also explicitly  demonstrate the quantisation
of the pumped charge, due to interference effects, when  we
increase the number of tunnel barriers ($n_b$) or equivalently
increase the number of dots ($n_d=n_b-1$) through which
the current is measured. We work with open quantum dots with
large transmissions and explicitly show how the charge
pumped through the device changes with the number of dots.
It is expected\cite{THOULESS,NIU} that for a fully spatially
periodic system, ($n_b$ or $n_d \rightarrow \infty$), the charge
pumped will be quantised. However, here we see that for reasonable
values of the barrier strengths and pumping strengths, 
(almost) quantisation occurs even with 4-6 dots. We also see that the pumped
charge, as a function of the number of dots, shows an interesting
step-like behaviour. The pumped charge increases with the 
number of barriers and then saturates near an integer.
Beyond that, it rises again with the number
of barriers, till it saturates at the next integer.
This tendency towards quantisation is also 
seen  as a function of the amplitude
of the pumping potential.
This clearly indicates the special stability that occurs when
an integer number of electrons are pumped through the system.
Note that this stability is purely quantum mechanical in
origin and {\it is not} due to interactions or CB 
physics, which allows electrons to be added only one by one
to the dot. Here, we are completely ignoring interaction effects
and there is no CB since the dots are well-coupled
to the leads.

Although we use $\delta$-function barriers for the explicit 
calculation, we expect the results to be robust to changing the
form of the barriers.
Following the work of Ref.\cite{NT}, we also expect these results
to be robust to weak disorder and to weak interactions. 

\begin{figure}[htb]
\begin{center}
\epsfig{figure=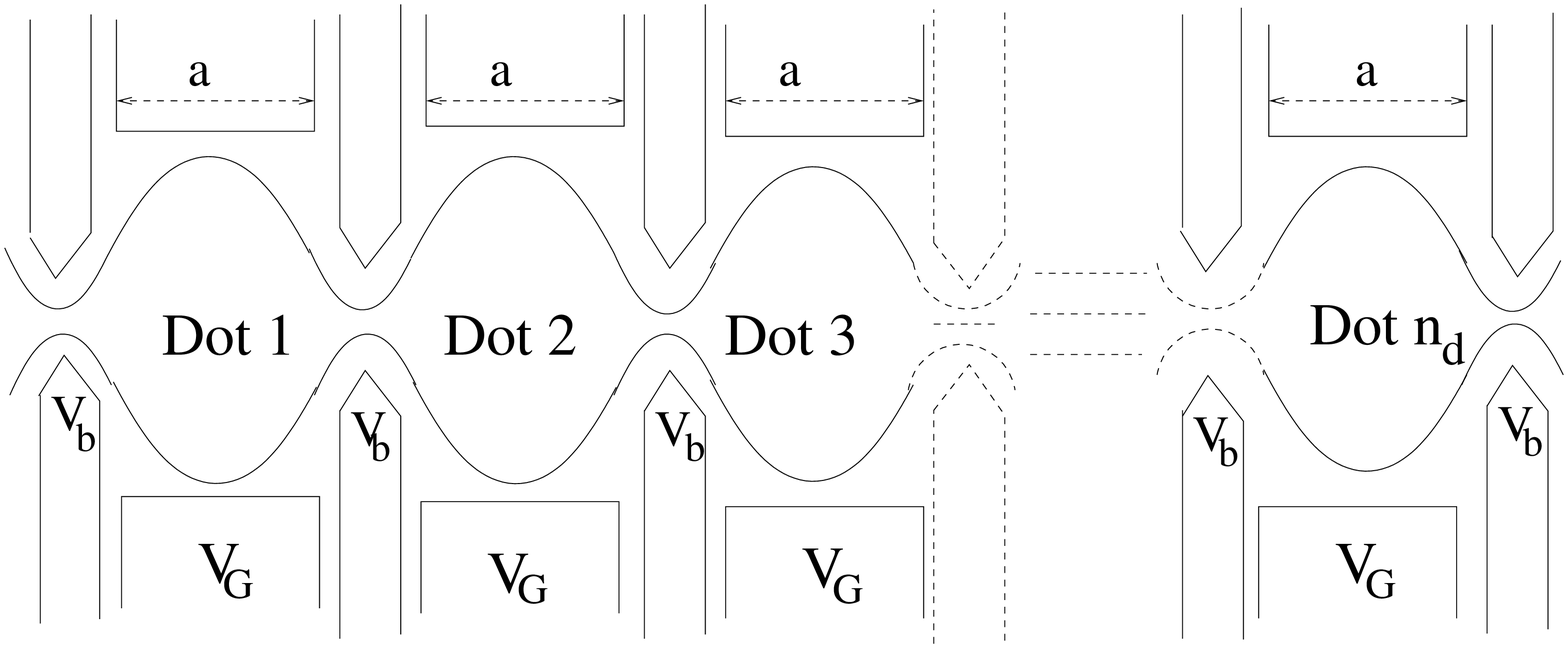,width=5.0cm}
\end{center}
\caption{Schematic diagram of a multiple dot
system ($n_d$ dots) defined on a two dimensional electron gas.  
The barriers forming the dots are denoted as $V_b$ and the gate 
voltages controlling the density in the dots are denoted as $V_G$.}
\end{figure}

We start with a system of coupled quantum dots as shown
in Fig. 1. The barriers forming the dot are 
periodically  modulated as 
\bea
V_i &\equiv& V_1 = V_0 + V_p \cos (\omega t), 
~i \le n_b/2  ~ {\rm for}~ n_b= {\rm even}, \nonumber \\ 
&&\quad i \le n_b/2+1 ~{\rm for}~ n_b={\rm odd}, \nonumber \\
V_i &\equiv& V_2= V_0 + V_p \cos (\omega t + \delta), ~{\rm for~
the ~remaining} .
\label{one}
\eea
Here $\omega$ is related to the time period as 
$\omega =\tau/2\pi$ and $\delta$ is the phase difference 
between the two time-varying potentials.
Such a potential breaks the parity symmetry
and allows the shape of 
all the dots to be varied. If the dots are coupled to the leads by a single
channel quantum point contact, then it is sufficient to treat the
dot within a one-dimensional effective Hamiltonian.
The width of the dots (effectively the width of the quantum well that
we use) is given by $a$.
We are mainly interested in the region where 
$V_0 \le E_F$ since we are in the resonant tunneling regime
and not in the CB regime.

The effective single channel $S$-matrix  for this system of $n_b$ barriers can 
be written as 
\beq
S = \left( \begin{array}{ll}
re^{i\theta} & t e^{i\phi} \\
t e^{i\phi} & r' e^{i\theta'} 
\end{array} \right)
\label{two}
\eeq
where the parameters $r,t,r',\theta,\theta'$ and $\phi$ are functions
of the Fermi energy $E_F$ and the amplitudes of the time-
varying potentials $V_i(t)$. Their explicit forms can be found, 
in terms of the 
parameters of a single well, (in the adiabatic limit),
by solving the
time-independent Schrodinger equation for the potential $V_i(t)$
given in Eq. \ref{one}, for each value of $t$.
The reflection amplitudes are not the same because the time-varying 
potentials explicitly violate parity. The potential also violates
time-reversal invariance. But since in the adiabatic approximation,
we are only interested in snapshots, at each value of the time, 
the Hamiltonian is time-reversal invariant and hence, the 
transmission amplitudes are the same for the 12 and 21 elements
in the $S$-matrix.

By the Brouwer formula\cite{BROUWER}, the charge pumped
can directly be computed from the parametric 
derivatives of the $S$-matrix. 
For a single channel, it  is given by 
\bea
Q &=& {e\over 2 \pi} \int_0^\tau dt Im {\large (}
{\partial S_{11}\over \partial V_1}S_{11}^* {\dot V_1}
+ {\partial S_{12}\over \partial V_1}S_{12}^* {\dot V_1}
\nonumber \\
&&~~~~~~~~~~\quad+{\partial S_{11}\over \partial V_2} S_{11}^*{\dot V_2}  
+ {\partial S_{12}\over \partial V_2} S_{12}^*{\dot V_2}{\large )}
\eea
where $S_{ij}$ denote the matrix elements of the $S$-matrix
and ${\dot V_1}$ and ${\dot V_2}$ are the time derivatives of
the $V_1,V_2$ given in Eq. \ref{one}.
For the form of the $S$-matrix given in Eq. \ref{two}, this is just
\beq
Q={e\over 2 \pi} \int_0^\tau dt (r^2 {\dot \theta} + t^2 {\dot \phi})~.
\label{pcharge}
\eeq 
Thus, the pumped charge is directly related to the amplitudes
and phases that appear in the scattering matrix.
Note that $Q$ can also be written in the form
of Eq. \ref{pcharge2}
where the first term is clearly quantised since $e^{i\theta}$ has to 
return to itself at the end of the period. So the only possible 
change in $\theta$ can be in integral multiples of $2\pi$.
The second term is the `dissipative' term which prevents
the perfect quantisation. It is easy to see the analogy
of Eq. \ref{pcharge2}  with Eq. 19 of Ref.\cite{ALEINER}.

The form  in Eq. \ref{pcharge} also  indicates 
that $Q$  is quantised whenever either $r$ or
$t$ is zero throughout the period.
When the Fermi energy lies in a gap, ($t=0,r=1$)
the charge is quantised. This is what happens for 
spatially  periodic potentials as discussed in Ref\cite{THOULESS,NIU}.
The charge is also quantised when there is total transmission
($t=1,r=0$) through almost the whole period.
This is essentially the case studied in Ref\cite{ENTIN},
where they find quantised charges whenever the pumping contour
encloses almost all of the resonance line.

In the rest of the letter, we 
compute the transmission and reflection coefficients, the phases
and the  quantised charges for various cases, and see that
the pumped charge is almost quantised even when the number
of barriers is quite small. We study the variation of $Q$
as a function of the number of barriers, as 
a function of $E_F$ and as a function of
the pumping amplitudes. We also study $Q$
as a function of the separation $a$ between the barriers and
as a function of the phase difference $\delta$. 

Strictly speaking, to remain within adiabatic approximation under which the 
Brouwer formula is derived, the energy level spacing in
the dots $\Delta$ has to be larger than the energy scale defined
by the frequency of the time-varying parameter $E_\omega = \hbar
\omega$. It is only under this approximation that the snapshot
picture of studying the static $S$-matrix for different time 
points within the period is valid. A better  approach to go beyond 
the adiabatic approximation\cite{WANG,BUTTIKER}
is to use the Floquet states.  However, here we use the
adiabatic approximation  even in the continuum limit ($E_F \ge V_0$)
where the energy levels are almost continuous and  $\Delta \rightarrow 0$, and  
a few energy levels cross the Fermi level within a period.
For sufficiently small $\omega$, we expect this approximation to
still yield qualitatively correct results.
 


\noindent $\bullet$  {\it Single dot case or $n_b=2$} :
Here, we compute the scattering matrix for two $\delta$-function 
barriers at a distance $a$ apart. Following Ref.\cite{WEI},
to obtain numerical results, we set $a=4$ and $\omega =1$.
We find, however, that our results are independent of $\omega$
and hence $\omega$ can be made as small as we wish.
Our energy units are set by $\hbar=2m=k_B=1$, where $k_B$ is the 
Boltzmann constant. So for $a=100 A^o$, which is a typical
value of the mean free path in $GaAs$, the energy unit in our
system of units is $E=5.6 meV$, which corresponds to a temperature
of $T=65 ^oK$.  With these units, we set $V_0=1$ and $V_p=0.4$.
We have also set the phase difference $\delta =\pi/2$ to obtain
the maximum pumped charge.
Here, we essentially reproduce the results of Wei et al\cite{WEI}.
We have also checked that the peak in the small pumped current, occurs
at the Fermi level, when one transmission maxima passes through the
Fermi level in one period. Also, since the transmission does not
become small (for a double-barrier system, the transmission does
not fall to zero after reaching a maximum), there is a large
dissipative term in the pumped charge, which explains why the
charge pumped is small. 

\begin{figure}[htb]
\begin{center}
\epsfig{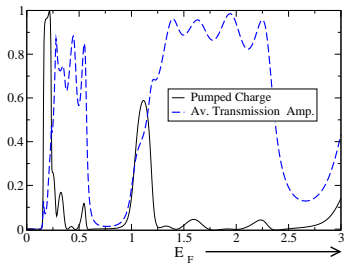}
\end{center}
\caption{$Q$ and the average transmission amplitude
per period versus $E_F$  for $n_b = 6$ barriers. We have
set $V_0=1$, $V_p=0.4$, $\omega =1$, $a=4$ and $\delta =\pi/2$.}
\end{figure}

\noindent $\bullet$ {\it Multiple dot case or $n_b > 2$} :
Here, we have computed the $S$-matrix, and obtained the transmission
and reflection coefficients, their phases and the pumped charge for
$n_b$ ranging from 3 to 14 ( 2 to 13 quantum dots).  
For each value of $n_b$,
the pumped charge and the average transmission 
amplitude (obtained by integrating the transmission amplitude
over one period and dividing by the period), 
is plotted as a function of the Fermi energy $E_F$,
between $E_F<V_0$ to $E>V_0$. This is shown in Fig. 2. for 
a typical case ($n_b=6$).
The pumped charge is a maximum just when the transmission rises
from zero.  Since $E_F \ll V_0$ is the
CB limit, which is not the limit we are studying,
our main focus is on the charge pumped when
$E\sim V_0$ (the second peak in Fig. 2). Peaks at higher values
of $E_F$ become progressively smaller. However, we often work with 
the first peak also for illustrative purposes, since we have
not included CB in our formalism.

\begin{figure}[htb]
\begin{center}
\epsfig{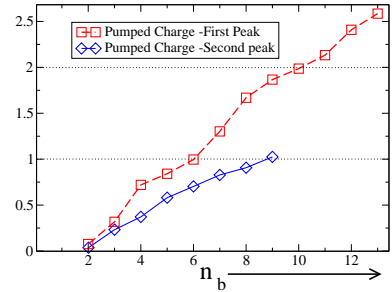}
\end{center}
\caption{$Q$ (in units of $e$)
as a function of the number of barriers
$n_b$. The top line is
for the first peak and the line below is for the second peak.
The parameter specifications are the same as for Fig.2.} 
\end{figure}

The magnitude of the (position of maximum
as a function of $E_F$)
pumped charge
as a function of $n_b$ is plotted in Fig. 3 for the first
and second peaks. Clearly, $Q$ 
increases with the number of barriers. The first peak
reaches quantisation with just $n_b =6$ whereas the second peak
requires $n_b=9$.
The most significant feature here is the step-like structure or
plateau structure near integer quantisation. 
(The quantisation for the first peak can be clearly seen,
at 1 for $n_b=6$, at 2 for $n_b=10$ and at 3 for $n_b=14$.
For the second
peak, we have only gone upto the value of one for $Q$,
since it is progressively more cumbersome to go to
larger number of barriers. But we have checked that the trend is the same.) 
For $n_b \rightarrow \infty$,
the system would be  spatially periodic and perfect quantisation would
have been  expected. However, 
what is interesting and unexpected is the formation of plateaux
for small values of $n_b$. 

\vspace{0.4cm}

\begin{figure}[htb]
\begin{center}
\epsfig{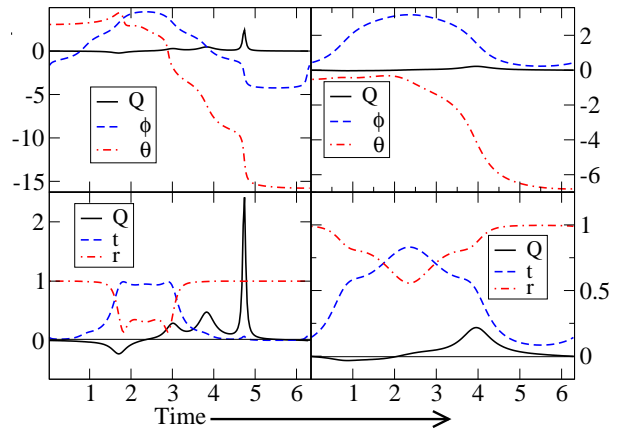}
\end{center}
\caption{Transmission and reflection amplitudes ($t$ and $r$) 
and phases ($\phi$ and $\theta$) and the pumped charge $Q$
as a function of time (for a single period $\tau=2\pi$)
for $n_b=10$ barriers (left panel) and $n_b=4$  barriers (right panel). 
$Q$ is measured in units of $e$ snd
$\phi$ and $\theta$ in  radians.}
\end{figure}


The correlation of the pumped charge with the values of the phases
and amplitudes of the reflection and transmission coefficients is shown in
Fig. 4  for a couple of typical cases.
In the top left panel, $Q$ is plotted  
along with the $\theta$ and $\phi$ as a function
of time for one full period,  for the 
system with $n_b=10$. For this system, we know from Fig. 3, that
the   total pumped charge is (almost) quantised 
at $Q=2$.  From the figure, we see that
$\phi$ comes back to itself at the
end of one period, but $\theta$ has a discontinuity.
It changes by $2 \times 2\pi$ in one period. Thus, the first
term in Eq. \ref{pcharge2} gives a factor of 2. The criterion for
quantisation is that the second term  should vanish.
The correlation of the quantisation with the (approximate) 
vanishing of the second term can also 
be seen by looking at the figure. In the left panel, we note that 
where the transmission 
amplitude $t$ is large, both $\theta$ and $\phi$  change very little  
and almost
symmetrically. On the other hand, where $\theta$  and $\phi$
change very  rapidly, the amplitude $t$ is very small.
In the  right panel, for contrast, we have studied
$n_b=4$  where
$Q$ is not quantised. We see that the change
in $\theta$  is  $2\pi$  so that the 
first term in Eq. \ref{pcharge2} gives 1. But here,
there are rapid  and non-symmetric changes in $\theta$ and $\phi$
when $t$ is large.  Hence, here, the second
term is non-zero and there is no quantisation of $Q$
near unity.
Features similar to the left  and
right panels  are  consistently seen whenever there is quantisation and
whenever there is no quantisation respectively.

The qualitative features described above
do not change when we change the ratio of $V_p$ to $V_0$.
In fact, as $V_p/V_0$ increases, we find that the value of
$Q$ and the tendency towards
quantisation increases. This is seen in Fig. 5.
We have also checked that $Q$ is periodic in the separation
$a$ (shown as inset in Fig. 6.). In the weak pumping limit, - $i.e.$, when the 
amplitude $V_p$  is suffiiciently small, we expect the charge
pumped to be proportional to $\sin \delta$\cite{BROUWER}, but
as $V_p$ increases, the sinusoidal shape is expected to
be distorted.  This feature is seen in 
Fig. 6. For $V_p \sim 0.05$, the dependence on $\delta$ is sinusoidal, 
but here $Q$ is  quite small. As $V_p$ increases, $Q$ increases, but 
there is also an 
increasing distortion of the sinusoidal shape.

\begin{figure}[htb]
\begin{center}
\epsfig{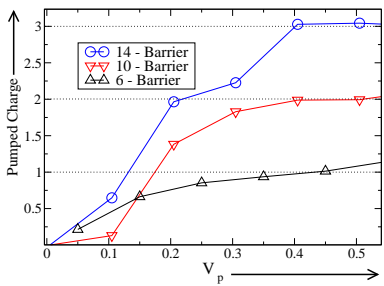}
\end{center}
\caption{$Q$ as a function of the amplitude $V_p$ of the 
time-varying potential for different $n_b$.}
\end{figure}


\begin{figure}[htb]
\begin{center}
\epsfig{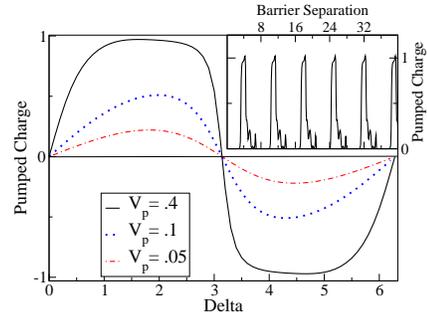}
\end{center}
\caption{Pumped charge as a function of the phase difference  $\delta$ 
for $n_b= 6$ barriers. The inset shows it as a function of the
separation $a$ between barriers.}
\end{figure}

To summarise, in this letter, we have shown that the pumped
charge shows an interesting step-like behaviour as we increase
the number of dots through which the charge is pumped and also
as we increase the pumping amplitude. We have 
derived and demonstrated the relation 
beween $Q$  and the
transmission and reflection coefficients of the effective $S$-matrix.
$Q$ is quantised whenever the 
contribution from the transmission amplitude vanishes -
$i.e.$, whenever there is no dissipation. The main point of
this letter is to emphasize that this can occur due to quantum 
interference, even for transmission through a few (4-6) dots
and not only for an infinite spatially periodic system.
The experimentally testable prediction here is that the
pumped charge {\it increases} as the number of barriers  through which
is is transmitted increases, and reaches quantisation with 
a few barriers.

\leftline{\bf Acknowledgments :} A. Banerjee 
would like to thank HRI for 
hospitality during the period of this work. 


\vskip -0.6 true cm

\end{document}